\documentclass[12pt]{article}
\begin{document}
\title{\bf  Teleparallel Killing Vectors of Spherically Symmetric Spacetimes}

\author{M. Sharif \thanks{msharif@math.pu.edu.pk} and Bushra
Majeed
\thanks{fmajeedpk@yahoo.com}\\
Department of Mathematics, University of the Punjab,\\
Quaid-e-Azam Campus, Lahore-54590, Pakistan.}

\date{}

\maketitle

\begin{abstract}
In this paper, Killing vectors of spherically spacetimes have been
evaluated in the context of teleparallel theory of gravitation.
Further, we investigate the Killing vectors of the Friedmann
metrics. It is found that for static spherically spacetimes the
number of Killing vectors turn out to be \emph{seven} while for the
Friedmann models, we obtain \emph{six} teleparallel Killing vectors.
The results are then compared with those of General Relativity. We
conclude that both of these descriptions of gravity do not provide
the consistent results in general. However, these results may
coincide under certain conditions for a particular spacetime.
\end{abstract}


\section{Introduction}

The concept of symmetry is much helpful to find the solution of
many problems in physics. The connection between symmetries and
conservation laws in physics has both fundamental significance and
great practical utility. The symmetries of curved spacetimes are
generated by Killing vectors (KVs) also called isometries. Some other
spacetime symmetries are Ricci collineations (RCs), curvature
collineations (CCs) and matter collineations (MCs). These
symmetries help to solve the Einstein field equations (EFEs) and
to classify their solutions. Thus symmetries play a significant
role in describing the geometry of a spacetime. Much work has been
done to explore these symmetries during the last two decades.
Petrov [1] was the first who considered the four-dimensional
spacetime to solve the Killing equations. Bokhari and Qadir [2,3]
gave a complete classification of static spherically symmetric
spacetimes. Ziad and Qadir [4,5] extended this work and obtained a
complete classification of non-static spherically symmetric
spacetimes.

Teleparallel theory of gravity (TPG) was introduced  by Einstein
in his unsuccessful attempt to unify electromagnetism and
gravitation [6]. This is considered as an alternative description
of gravitation which corresponds to a gauge theory for the
translation group [7] based on Weitzenb\"{o}ck geometry [8]. Here
the non-zero torsion plays the role of force [9] and curvature
tensor vanishes identically. In this approach, there are no
geodesics and the gravitational interaction is described by force
equations similar to the Lorentz force equations of
electrodynamics [10]. Thus we can say that the gravitational
interaction can be described in terms of curvature as is usually
done in GR or in terms of torsion as in TPG.

In this theory, tetrad plays the role of basic entity instead of
the metric tensor as in the case of GR. The tetrad formulation of
gravitation was considered by M{\o}ller with attempts to define
the energy of the gravitational field. In 1961, M{\o}ller revived
Einstein's idea of unification of gravitation and electromagnetism
and presented a theory which based on tetrad field instead of
metric tensor [11]. Later, Pellegrini and Plebanski [12] found a
Lagrangian formulation for absolute parallelism. Hayashi and
Nakano [13] independently worked to formulate the gauge theory of
the spacetime translation group. Hayashi [14] also pointed out the
connection between the gauge theory of spacetime translation group
and absolute parallelism. During the last two decades, much
attention has been given to analyze the basic solutions of GR with
TPG and comparing the results. There is a great literature
available [15-24] on the study of TP versions of the exact
solutions of GR.

Recently, Sharif and Jamil [25,26] have found the TP versions of
the Friedmann models, Lewis-Papapetrou spacetimes and stationary
axisymmetric solutions of the Einstein-Maxwell field equations.
They have explored the energy-momentum distribution of these
solutions [27] by using certain energy-momentum density developed
from M{\o}ller's tetrad theory. Further, they have evaluated the
energy contents of static axially symmetric spacetimes [28] by
using TP version of four prescriptions, namely, Einstein,
Landau-Lifshitz, Bergmann-Thomson and prescription developed from
M{\o}ller's tetrad theory. In a recent paper, the same authors
[29] introduced the idea of symmetry of a spacetime with torsion
only. For this purpose, they defined the TP version of Lie
derivative and found the corresponding Killing equations. They
evaluated the TP KVs of the Einstein universe.

This paper extends the above work by evaluating the TP KVs of static
spherically symmetric and the Friedmann spacetimes. The results are
compared with those available in GR. The layout of the paper is as
follows. In section \textbf{2}, we shall give brief review of
fundamental concepts of TPG. Section \textbf{3} is devoted to
explore the KVs of static spherically symmetric spacetimes. In
section \textbf{4}, the TP KVs of the Friedmann metrics have been
investigated. The last section contains discussion and conclusion of
the results obtained.

\section{A Brief Review of Teleparallel Theory of Gravity}

In TPG, the gravitational field is described by the tetrad field
${h^a}_{\mu}$ which has been used to define the Weitzenb\"{o}ck
connection [10] as
\begin{equation}\label{1}
{{\Gamma^\theta}_{\mu\nu}} =
{{h_a}^{\theta}}\partial_{\nu}{{h^a}_{\mu}},
\end{equation}
with respect to which tetrad is parallel:
\begin{equation}\label{2}
\nabla_{\nu}{{h^a}_{\mu}}\equiv{{h^a}_{\mu,\nu}}
-{{\Gamma^\theta}_{\mu\nu}}{{h^a}_{\theta}}=0.
\end{equation}
This is called the absolute parallelism condition, i.e., tetrad are
parallely transported in the Weitzenb\"{o}ck spacetime. Greek
alphabets $(\mu, \nu,\rho, ...=0,1,2,3)$ denote spacetime indices
and Latin alphabets $(a,b,c,...=0,1,2,3)$ represent tangent space
indices. Spacetime indices and tangent space indices can be replaced
by tetrad field satisfying
\begin{equation}\label{3}
{{h^a}_\mu}{{h_a}^\nu}={{\delta_\mu}^\nu},\quad {{h^a}
_\mu}{{h_b}^\mu}={{\delta^a}_b}.
\end{equation}
The presence of non-trivial tetrad field induces both Riemannian
and TP structures on a spacetime and gives rise to the Riemannian
metric as a by product [10], i.e.,
\begin{equation}\label{4}
g_{\mu\nu} = \eta_{ab}{{h^a}_{\mu}}{{h^b}_{\nu}}.
\end{equation}
The torsion of the Weitzenb\"{o}ck connection is defined as
\begin{equation}\label{5}
{{T^\theta}_{\mu\nu}}={{\Gamma^\theta}_{\nu\mu}}
-{{\Gamma^\theta}_{\mu\nu}}.
\end{equation}
The relationship between the Weitzenb\"{o}ck and Levi-Civita
connections is
\begin{equation}\label{6}
{{\Gamma^\theta}_{\mu\nu}}={{\Gamma^{0}}^\theta}_{\mu\nu}+
{{K^\theta}_{\mu\nu}},
\end{equation}
where
\begin{equation}\label{7}
{{K^\theta}_{\mu\nu}}=\frac{1}{2}[T_{\mu~~\nu}^{~\theta}+
T_{\nu~~\mu}^{~\theta}-{{T^\theta}_{\mu\nu}}]
\end{equation}
is a tensor quantity known as \textbf{contortion} tensor and
${{\Gamma^{0}}^\theta}_{\mu\nu}$ is the Levi-Civita connection of GR
\begin{equation}\label{8}
{{\Gamma^{0}}^\theta}_{\mu\nu}
=\frac{1}{2}g^{\theta\sigma}(\partial_{\mu}g_{\sigma\nu}
+\partial_{\nu}g_{\sigma\mu}+\partial_{\sigma}g_{\mu\nu}).
\end{equation}
The TP Lie derivative of a covariant tensor of rank $2$ along a
vector field $\xi$ is defined [29] as
\begin{equation}\label{9}
(\pounds_{\xi}A)_{\mu\nu}=\xi^{\rho}\nabla_{\rho}A_{\mu\nu}
+(\nabla_{\mu}\xi^{\rho})A_{\rho\nu}
+(\nabla_{\nu}\xi^{\rho})A_{\mu\rho},
\end{equation}
where $\nabla_{\rho}$ represents TP covariant derivative given as
\begin{equation}\label{10}
\nabla_{\rho}A_{\mu\nu} = A_{\mu\nu},_{\rho}
-{{\Gamma^\theta}_{\rho\nu}}A_{\mu\theta}
-{{\Gamma^\theta}_{\rho\mu}}A_{\theta\nu}.
\end{equation}
In terms of torsion tensor, this can be written as
\begin{equation}\label{11}
(\pounds_{\xi}A)_{\mu\nu}=\xi^{\rho}A_{\mu\nu},_{\rho}+
A_{\rho\nu}\xi^{\rho},_{\mu}+A_{\mu\rho}\xi^{\rho},_{\nu}+
\xi^{\rho}(A_{\theta\nu}{{T^\theta}_{\mu\rho}}+
A_{\mu\theta}{{T^\theta}_{\nu\rho}}).
\end{equation}
Similarly, for a contravariant tensor of rank $2$, the TP Lie
derivative turns out to be
\begin{equation}\label{12}
(\pounds_{\xi}A)^{\mu\nu}=\xi^{\rho}A^{\mu\nu},_{\rho}
-A^{\rho\nu}\xi^{\mu},_{\rho}-A^{\mu\rho}\xi^{\nu},_{\rho}
-\xi^{\rho}(A^{\theta\nu}{{T^\mu}_{\theta\rho}}
+A^{\mu\theta}{{T^\nu}_{\theta\rho}}).
\end{equation}
The extension of this definition to a mixed tensor of rank $r+ s$
is given as
\begin{eqnarray}\label{13}
{(\pounds_{\xi}A)^{\rho...\sigma}}_{\mu...\nu}&=&
\xi^{\alpha}{A^{\rho...\sigma}}_{\mu...\nu},_{\alpha}+
{A^{\rho...\sigma}}_{\alpha...\nu}\xi^{\alpha},_{\mu}+
{A^{\rho...\sigma}}_{\mu...\alpha}\xi^{\alpha},_{\nu}\nonumber\\
&-&{A^{\alpha...\sigma}}_{\mu...\nu}\xi^{\rho},_{\alpha}-...-
{A^{\rho...\alpha}}_{\mu...\nu}\xi^{\sigma},_{\alpha}\nonumber\\
&+&\xi^{\alpha}({A^{\rho...\sigma}}_{\beta...\nu}{T^\beta}_{\mu
\alpha}
+...+{A^{\rho...\sigma}}_{\mu...\beta}{T^\beta}_{\nu\alpha} \nonumber\\
&-&{A^{\beta...\sigma}}_{\mu...\nu}{T^\rho}_{\beta\alpha}-...-
{A^{\rho...\beta}}_{\mu...\nu}{T^\sigma}_{\beta\alpha}),
\end{eqnarray}
where $|{\rho...\sigma}|=r$ and  $|{\mu...\nu}|=s$.

The Killing equations in TPG are defined as
\begin{equation}\label{14}
\pounds_{\xi}g_{\mu\nu}=0.
\end{equation}
In component form, it can be expressed as
\begin{equation}\label{15}
g_{\mu\nu,\rho}\xi^{\rho}+g_{\rho\nu}\xi^{\rho},_{\mu}
+g_{\mu\rho}\xi^{\rho},_{\nu}
+\xi^{\rho}(g_{\theta\nu}{{T^\theta}_{\mu\rho}}
+g_{\mu\theta}{{T^\theta}_{\nu\rho}})=0.
\end{equation}

\section{TP Killing Vectors of Static Spherically Symmetric Spacetimes}

The most general static spherically symmetric spacetime is given
[30] as
\begin{equation}\label{17}
ds^2
=e^{\nu(r)}dt^2-e^{\lambda(r)}dr^2-r^2d\theta^2-r^2\sin^2\theta
d\phi^2.
\end{equation}
Using the procedure [29], the non-vanishing components of the
Weitzenb\"{o}ck connection are of the form
\begin{eqnarray}\label{23}
{\Gamma^0}_{01}&=&[\ln\sqrt{g_{00}}]_{,r},\quad
{\Gamma^2}_{12}=\frac{\sqrt{-g_{11}}}{r}= {\Gamma^3}_{13},\quad
{\Gamma^1}_{11}=[\ln\sqrt{-g_{11}}]_{,r},\nonumber\\
{\Gamma^2}_{21}&=&\frac{1}{r}={\Gamma^3}_{31},\quad
{\Gamma^1}_{22}=\frac{-r}{\sqrt{-g_{11}}},\quad{\Gamma^3}_{23}
=\cot\theta={\Gamma^3}_{32},\nonumber\\
{\Gamma^2}_{33}&=&-\sin\theta\cos\theta,\quad {\Gamma^1}_{33}
={\Gamma^1}_{22}\sin^2\theta.
\end{eqnarray}
The only non-zero components of the torsion tensor are
\begin{eqnarray}\label{24}
{T^0}_{01}=-[\ln\sqrt{g_{00}}]_{,r},\quad
{T^2}_{21}=-\frac{1}{r}(1-\sqrt{-g_{11}})={T^3}_{31}.
\end{eqnarray}
Using these values of the components of torsion tensor in
Eq.(\ref{15}), we get the following system of ten partial
differential equations
\begin{eqnarray}\label{25}
\xi^0=A(r,\theta,\phi),&&\\\label{26}
e^\lambda{\xi^1}_{,0}-e^\nu{\xi^0}_{,1}
-\frac{\nu'}{2}e^\nu\xi^0=0,&&\\\label{27}
r^2{\xi^2}_{,0}-e^\nu{\xi^0}_{,2}=0,&&\\\label{28}
r^2\sin^2{\xi^3}_{,0}-e^\nu{\xi^0}_{,3}=0,&&
\\\label{29}\lambda'\xi^1+2
{\xi^1}_{,1}=0,&&\\\label{30}
{\xi^2}_{,1}+\frac{e^\lambda}{r^2}{\xi^1}_{,2}
+\frac{1-\sqrt{e^\lambda}}{r}\xi^2=0,&&
\\\label{31}{\xi^3}_{,1}+\frac{e^\lambda}{r^2\sin^2\theta}{\xi^1}_{,3}
+\frac{1-\sqrt{e^\lambda}}{r}\xi^3=0,&&
\\\label{32}\xi^1+\frac{r}{\sqrt{e^\lambda}}{\xi^2}_{,2}=0,&&
\\\label{33}{\xi^2}_{,3}+ \sin^2 \theta
{\xi^3}_{,2}=0,&&\\\label{34}\sqrt{e^\lambda}\xi^1 + r \cot\theta
\xi^2 + r {\xi^3}_{,3}=0,&&
\end{eqnarray}
where prime represents derivative with respect to $r$.

Solving Eqs.(\ref{29}), (\ref{30}) and (\ref{32}) simultaneously,
we get
\begin{eqnarray}\label{35}
\xi^1&=& e^{-\frac{\lambda}{2}}\{B_1(t,\phi)\cos\theta+B_2(t,\phi)\sin\theta\},\\
\label{36}\xi^2&=&-\frac{1}{r}\{B_1(t,\phi)\sin\theta-B_2(t,\phi)\cos\theta\}+
C(t,\phi)e^{-F(r)},
\end{eqnarray}
where
\begin{equation}\label{37}
F(r)=\int\frac{1-e^{\frac{\lambda}{2}}}{r}dr.
\end{equation}
Substituting this value of $\xi^2$ in Eq.(\ref{33}), we have
\begin{eqnarray}\label{38}
\xi^3&=&\frac{1}{r}\{{B_{1,3}}(t,\phi)\ln|\csc\theta-\cot\theta|
+{B_{2,3}}(t,\phi)\csc\theta\} \nonumber\\
&-&{C_{,3}}(t,\phi)e^{-F(r)}\cot\theta+D(t,r,\phi).
\end{eqnarray}
Using these values in the remaining equations, it follows that
\begin{eqnarray}\label{39}
\xi^0&=&e^{-\frac{\nu}{2}}c_0,\nonumber\\
\xi^1&=&e^{-\frac{\lambda}{2}}\{c_1\cos\theta
+(c_2\cos\phi+c_3\sin\phi)\sin\theta\},\nonumber\\
\xi^2&=&-\frac{1}{r}\{c_1\sin\theta-(c_2\cos\phi+c_3\sin\phi)\cos\theta\}
+e^{-F(r)}(c_4\cos\phi+c_5\sin\phi),\nonumber\\
\xi^3&=&\frac{1}{r}\{(c_3\cos\phi-c_2\sin\phi)\csc\theta\} \nonumber\\
&-&e^{-F(r)}\cot\theta(c_4\sin\phi-c_5\cos\phi)+c_6e^{-F(r)}.
\end{eqnarray}
Thus we obtain seven TP KVs of static spherically symmetric
spacetimes given as
\begin{eqnarray}\label{40}
\xi_{(0)}&=&e^{-\frac{\nu}{2}}\frac{\partial}{\partial t},\nonumber\\
\xi_{(1)}&=& e^{-F(r)} \frac{\partial}{\partial \phi},\nonumber\\
\xi_{(2)}&=& e^{-F(r)}(\cos\phi \frac{\partial}{\partial
\theta}-\cot\theta\sin\phi \frac{\partial}{\partial
\phi}),\nonumber\\
\xi_{(3)}&=& e^{-F(r)}(\sin\phi \frac{\partial}{\partial
\theta}+\cot\theta\cos\phi \frac{\partial}{\partial
\phi}),\nonumber\\
\xi_{(4)}&=& e^{-\frac{\lambda}{2}} \cos\theta
\frac{\partial}{\partial r}-\frac{1}{r}\sin \theta
\frac{\partial}{\partial \theta},\nonumber\\
\xi_{(5)}&=& e^{-\frac{\lambda}{2}} \sin\theta \cos\phi
\frac{\partial}{\partial r}+\frac{1}{r}\cos\phi \cos\theta
\frac{\partial}{\partial \theta}-\frac{1}{r}\sin\phi \csc\theta
\frac{\partial}{\partial \phi},\nonumber\\
\xi_{(6)}&=& e^{-\frac{\lambda}{2}} \sin\theta \sin\phi
\frac{\partial}{\partial r}+\frac{1}{r}\sin\phi \cos\theta
\frac{\partial}{\partial \theta}+\frac{1}{r}\cos\phi \csc\theta
\frac{\partial}{\partial \phi}.
\end{eqnarray}

\section{TP Killing Vectors of the Friedmann Models}

The metric representing the FRW models is given [31] as
\begin{equation}\label{64}
ds^2=dt^2-a^2(t)[d\chi^2+{f_k}^2(\chi) (d\theta^2+\sin^2\theta
d\phi^2)],
\end{equation}
where
\begin{eqnarray}\label{65}
f_k(\chi)&=&\sinh\chi,\quad k=-1,\nonumber\\
&=&\chi,\quad\quad\quad k=0,\nonumber\\
&=&\sin\chi,\quad k=+1,
\end{eqnarray}
The non-zero components of the Weitzenb$\ddot{o}$ck connection are
\begin{eqnarray}\label{68}
{\Gamma^1}_{10}&=&{\Gamma^2}_{20}={\Gamma^3}_{30}={\frac{\dot{a}}{a}},\nonumber\\
{\Gamma^1}_{22}&=&-f_k,\quad\
{\Gamma^1}_{33}={\Gamma^1}_{22}\sin^2\theta,\nonumber\\
{\Gamma^2}_{12}&=&{\Gamma^3}_{13}={\frac{1}{f_k}}, \quad\
{\Gamma^2}_{21}={\Gamma^3}_{31}={\Gamma^2}_{12}
f'_k ,\nonumber\\
{\Gamma^2}_{33}&=&-\sin{\theta}\cos\theta, \quad\
{\Gamma^3}_{23}=\cot\theta={\Gamma^3}_{32},
\end{eqnarray}
where dot and prime denote the derivatives with respect to $t$ and
$\chi$ respectively. Consequently, the non-vanishing components of
the torsion tensor are
\begin{eqnarray}\label{69}
{T^1}_{10}&=&{T^2}_{20}={T^3}_{30}=- {T^1}_{01}=- {T^2}_{02}=-
{T^3}_{03}=-\frac{\dot{a}}{a},\nonumber\\
{T^2}_{21}&=&{T^3}_{31}=-{T^2}_{12}=
-{T^3}_{13}=\frac{1}{f_k}(1-f'_k).
\end{eqnarray}
In component form the TP Killing equations for this metric turn
out to be
\begin{eqnarray}\label{70}
\xi^0=A(\chi,\theta,\phi),&&\\\label{71}
\xi^1=B(t,\theta,\phi),&&\\\label{72}
{\xi^2}_{,3}+\sin^2\theta{\xi^3}_{,2}=0,&&\\\label{73}
a^2{\xi^1}_{,0}-{\xi^0}_{,1}+a\dot{a}\xi^1=0,&&\\\label{74}
\dot{a}f_k\xi^0+af_k{\xi^2}_{,2}+a\xi^1=0,&&\\\label{75}
{f_k}^2{\xi^2}_{,1}+{\xi^1}_{,2}-f_k\{1-f'_k\}
\xi^2=0,&&\\\label{76}
{f_k}^2\sin^2\theta{\xi^3}_{,1}+{\xi^1}_{,3}
-{f_k}\{1-f'_k\}\sin^2\theta\xi^3=0,&&\\\label{77}
\dot{a}{f_k}\xi^0+a\xi^1+a{f_k}\cot\theta\xi^2
+a{f_k}{\xi^3}_{,3}=0,&&\\\label{78}
a^2{f_k}^2{\xi^2}_{,0}-{\xi^0}_{,2}+a\dot{a}{f_k}^2\xi^2=0,&&
\\\label{79}
{a^2}{f_k}^2\sin^2\theta{\xi^3}_{,0}-{\xi^0}_{,3}
+a\dot{a}{f_k}^2{\sin^2\theta}\xi^3=0.&&
\end{eqnarray}

Solving Eqs.(\ref{74}), (\ref{75}) and (\ref{78}) along with
Eqs.(\ref{70}) and (\ref{71}), we get $a=c_1t+c_2$ and
\begin{eqnarray}\label{80}
\xi^0&=&e^{F(\chi)}\{A_1(\phi)\theta+A_2(\phi)\},\nonumber\\
\xi^1&=&\frac{1}{a}\{B_1{(\phi)}\cos\theta+B_2{(\phi)}\sin\theta\},\nonumber\\
\xi^2&=&-\frac{\dot{a}}{a}e^{F(\chi)}\{A_1(\phi)\frac{\theta^2}{2}
+A_2(\phi)\theta\}\nonumber\\
&-&\frac{1}{af_k}\{B_1{(\phi)}\sin\theta-B_2{(\phi)}\cos\theta\}+C(t,\chi,\phi),
\end{eqnarray}
where
\begin{equation}\label{81}
{F(\chi)}=\int\frac{1-f'_k}{f_k}d\chi.
\end{equation}
Now solving Eqs.(\ref{72}) and (\ref{77}) along with Eq.(\ref{80}),
we have
\begin{eqnarray}\label{82}
\xi^0&=&e^{F(\chi)}(c_3\phi+c_4)\theta+c_5\phi+c_6,\nonumber\\
\xi^1&=&\frac{1}{a}\{(c_7\phi+c_8)\cos\theta+(c_9\cos\phi
+c_{10}\sin\phi)\sin\theta\},\nonumber\\
\xi^2&=&-\frac{\dot{a}}{a}e^{F(\chi)}\{(c_3\phi
+c_4)\frac{\theta^2}{2}+(c_5\phi+c_6)\theta\}\nonumber\\
&-&\frac{1}{af_k}\{(c_7\phi+c_8)\sin\theta-(c_9\cos\phi
+c_{10}\sin\phi)\cos\theta\}\nonumber\\
&+&C_1(t,\chi)\cos\phi+C_2(t,\chi)\sin\phi,\nonumber\\
\xi^3&=&-\frac{\dot{a}}{a}e^{F(\chi)}\{(c_3\frac{\phi^2}{2}
+c_4\phi)\theta+(c_5\frac{\phi^2}{2}+c_6\phi)\}\nonumber\\
&-&\frac{1}{af_k}\{(c_9\sin\phi-c_{10}\cos\phi)\sin\theta\}\nonumber\\
&+&\cot\theta\frac{\dot{a}}{a}e^{F(\chi)}\{(c_3\frac{\phi^2}{\phi}+c_4\phi)\frac{\theta^2}{2}
+(c_5\frac{\phi^2}{2}+c_6\phi)\theta\}\nonumber\\
&-&\frac{\cot\theta}{af_k}\{(c_9\sin\phi-c_{10}\cos\phi)\cos\theta\}
\nonumber\\&-&\cot\theta\{C_1(t,\chi)\sin\phi-C_2(t,\chi)\cos\phi\}+D(t,\chi,\theta),
\end{eqnarray}
Using these values of $\xi^0,~\xi^1,~\xi^2$ and $\xi^3$ in the
remaining equations, we obtain
\begin{eqnarray}\label{83}
\xi^0&=&0,\nonumber\\
\xi^1&=&\frac{1}{a}\{c_1\cos\theta+(c_2\cos\phi+c_3\sin\phi)\sin\theta\},\nonumber\\
\xi^2&=&-\frac{1}{af_k}\{c_1\sin\theta-(c_2\cos\phi+c_3\sin\phi)\cos\theta\}
+\frac{e^{F(\chi)}}{a}\{c_4\cos\phi+c_5\sin\phi\},\nonumber\\
\xi^3&=&-\frac{1}{af_k}\{(c_2\sin\phi-c_3\cos\phi)\sin\theta\}
-\frac{\cot\theta}{af_k}\{(c_2\sin\phi-c_3\cos\phi)\cos\theta\}\nonumber\\
&-&\frac{e^{F(\chi)}\cot\theta}{a}\{c_4\sin\phi-c_5\cos\phi\}+c_6\frac{e^{F(\chi)}}{a}.
\end{eqnarray}
This leads to six TP KVs of FRW spacetimes
\begin{eqnarray}\label{84}
\xi_{(0)}&=&\frac{e^{F(\chi)}}{a}\frac{\partial}{\partial
\phi},\nonumber\\
\xi_{(1)}&=&\frac{e^{F(\chi)}}{a}\{\sin\phi\frac{\partial}{\partial
\theta}+\cot\theta\cos\phi\frac{\partial}{\partial
\phi}\},\nonumber\\
\xi_{(2)}&=&\frac{e^{F(\chi)}}{a}\{\cos\phi\frac{\partial}{\partial
\theta}-\cot\theta\sin\phi\frac{\partial}{\partial
\phi}\},\nonumber\\
\xi_{(3)}&=&\frac{1}{a}\{\cos\theta\frac{\partial}{\partial
\chi}-\frac{1}{f_k}\sin\theta\frac{\partial}{\partial
\theta}\},\nonumber\\
\xi_{(4)}&=&\frac{1}{a}\{(\sin\theta\frac{\partial}{\partial \chi}
+\frac{1}{f_k}\cos\theta\frac{\partial}{\partial
\theta})\cos\phi-\frac{1}{f_k}(\sin\theta+\cot\theta\cos\theta)
\sin\phi\frac{\partial}{\partial\phi}\},\nonumber\\
\xi_{(5)}&=&\frac{1}{a}\{(\sin\theta\frac{\partial}{\partial
\chi}+\frac{1}{f_k}\cos\theta\frac{\partial}{\partial
\theta})\sin\phi+\frac{1}{f_k}(\sin\theta+\cot\theta\cos\theta)
\cos\phi\frac{\partial}{\partial\phi}\}.\nonumber\\
\end{eqnarray}

\section{Summary and Discussion}

This paper is devoted to investigate the symmetries in the
teleparallel theory of gravitation. We have evaluated TP KVs for
static spherically symmetric spacetimes and the Friedmann metrics.
We find that there exist seven TP KVs of static spherically
symmetric spacetimes while there are four KVs in GR. If we compare
KVs of TPG and GR, it follows that $\xi_{(0)}$ and
$\xi_{(\rho)}~(\rho=1,2,3)$ in TPG are multiple of the corresponding
KVs in GR by $e^{-\frac{\nu}{2}}$ and $e^{-F(r)}$ respectively. The
first four TP KVs coincide to the four KVs of GR (as given in
appendix \textbf{A}) for $\nu=0$ and $F(r)=0$, i.e., for
$\lambda=0$. These conditions reduce the metric (\ref{17}) to
Minkowski spacetime in spherical polar coordinates. On the other
hand, under these assumptions, Eq.(\ref{24}) implies that all
components of the torsion tensor will become zero and hence the TP
Killing equations will reduce to the Killing equations of GR. This
yields the trivial four KVs of the spherical symmetry. The remaining
three TP KVs are due to the non-vanishing components of the torsion
tensor involving in the TP Killing equations. In view of these
results, it can be conjectured that:\\
\textit{In TPG, the number of independent KVs are restricted to
seven for all static spherically symmetric spacetimes}.

We have also explored TP KVs of the Friedman models. It is found
that these are six TP KVs for each model similar to GR [32].
However, the comparison of these results with those in GR shows
that, for $k=0$, the first four TP KVs $\xi_{(\rho)}$
($\rho=0,1,2,3$) are multiple of the corresponding KVs of GR by
$\frac{1}{a}$ while the remaining KVs are different. This difference
occurs because of the presence of the torsion components in TP
Killing equations. For $a(t)=1$, the first four KVs are the same in
both GR (as given in appendix A) and TPG which leads to Minkowski
spacetime. In TPG, the value of scale factor is found to be
$a(t)=c_1t+c_2$ which exactly matches with the value found [30] in
GR for $c_1=\frac{1}{b}$ and $c_2=0$. For both open and closed FRW
models, the KVs do not match in both the theories. This difference
is due to the same reason and hence the Lie algebra of these TP KVs
is not closed.

We conclude that the results of KVs do not coincide for the two
theories. It seems that the reason is the non-vanishing components
of torsion. It would be interesting to explore the compatibility of
the two theories TPG and GR and interpret exactly the TP KVs
corresponding to GR.

\renewcommand{\theequation}{A\arabic{equation}}
\setcounter{equation}{0}
\section*{Appendix A}

\section*{Killing Vectors of Static Spherically Symmetric Spacetimes in GR}

Killing vectors for static spherically symmetric spacetimes in GR
are
\begin{eqnarray}\label{E:5.0.1}
\xi_{(0)}&=&\frac{\partial}{\partial t},\nonumber\\
\xi_{(1)}&=& \frac{\partial}{\partial \phi},\nonumber\\
\xi_{(2)}&=& \cos\phi \frac{\partial}{\partial
\theta}-\cot\theta\sin\phi \frac{\partial}{\partial
\phi},\nonumber\\
\xi_{(3)}&=& \sin\phi \frac{\partial}{\partial
\theta}+\cot\theta\cos\phi \frac{\partial}{\partial \phi}.
\end{eqnarray}

\section*{Killing Vectors of Friedmann Spacetimes in GR}

Linearly independent KVs associated with the FRW models are [43]:\\
For $k=0$
\begin{eqnarray}\label{E:5.0.1}
\xi_{(0)}&=&\frac{\partial}{\partial
\phi},\nonumber\\
\xi_{(1)}&=&\sin\phi\frac{\partial}{\partial
\theta}+\cot\theta\cos\phi\frac{\partial}{\partial
\phi},\nonumber\\
\xi_{(2)}&=&\cos\phi\frac{\partial}{\partial
\theta}-\cot\theta\sin\phi\frac{\partial}{\partial
\phi},\nonumber\\
\xi_{(3)}&=&(\cos\theta\frac{\partial}{\partial
\chi}-\frac{1}{\chi}\sin\theta\frac{\partial}{\partial \theta}),\nonumber\\
\xi_{(4)}&=&(\sin\theta\frac{\partial}{\partial
\chi}+\frac{1}{\chi}\cos\theta\frac{\partial}{\partial
\theta})\cos\phi-\frac{1}{\chi}\csc\theta\sin\phi\frac{\partial}{\partial
\phi},\nonumber\\
\xi_{(5)}&=&(\sin\theta\frac{\partial}{\partial
\chi}+\frac{1}{\chi}\cos\theta\frac{\partial}{\partial
\theta})\sin\phi+\frac{1}{\chi}\csc\theta\cos\phi\frac{\partial}{\partial
\phi}.
\end{eqnarray}
For $k=1$
\begin{eqnarray}
\xi_{(0)}&=&\frac{\partial}{\partial
\phi},\nonumber\\
\xi_{(1)}&=&\sin\phi\frac{\partial}{\partial
\theta}+\cot\theta\cos\phi\frac{\partial}{\partial
\phi},\nonumber\\
\xi_{(2)}&=&\cos\phi\frac{\partial}{\partial
\theta}-\cot\theta\sin\phi\frac{\partial}{\partial
\phi},\nonumber\\
\xi_{(3)}&=&(\cos\theta\frac{\partial}{\partial
\chi}-\cot\chi\sin\theta\frac{\partial}{\partial \theta}),\nonumber\\
\xi_{(4)}&=&(\sin\theta\frac{\partial}{\partial
\chi}+\cot\chi\cos\theta\frac{\partial}{\partial
\theta})\cos\phi-\cot\chi\csc\theta\sin\phi\frac{\partial}{\partial
\phi},\nonumber\\
\xi_{(5)}&=&(\sin\theta\frac{\partial}{\partial
\chi}+\cot\chi\cos\theta\frac{\partial}{\partial
\theta})\sin\phi+\cot\chi\csc\theta\cos\phi\frac{\partial}{\partial
\phi}.
\end{eqnarray}
For $k=-1$
\begin{eqnarray}
\xi_{(0)}&=&\frac{\partial}{\partial
\phi},\nonumber\\
\xi_{(1)}&=&\sin\phi\frac{\partial}{\partial
\theta}+\cot\theta\cos\phi\frac{\partial}{\partial
\phi},\nonumber\\
\xi_{(2)}&=&\cos\phi\frac{\partial}{\partial
\theta}-\cot\theta\sin\phi\frac{\partial}{\partial
\phi},\nonumber\\
\xi_{(3)}&=&(\cos\theta\frac{\partial}{\partial
\chi}-\coth\chi\sin\theta\frac{\partial}{\partial \theta}),\nonumber\\
\xi_{(4)}&=&(\sin\theta\frac{\partial}{\partial
\chi}+\coth\chi\cos\theta\frac{\partial}{\partial
\theta})\cos\phi-\coth\chi\csc\theta\sin\phi\frac{\partial}{\partial
\phi},\nonumber\\
\xi_{(5)}&=&(\sin\theta\frac{\partial}{\partial
\chi}+\coth\chi\cos\theta\frac{\partial}{\partial
\theta})\sin\phi+\coth\chi\csc\theta\cos\phi\frac{\partial}{\partial
\phi}.
\end{eqnarray}
{\bf Acknowledgment:} We would like to thank Mr. Jamil Amir for
fruitful discussions during this work.\\

{\bf \large References}

\begin{description}

\item{[1]} Petrov, A.Z.: \textit{Einstein Spaces} (Pergamon, Oxford University Press, 1989).

\item{[2]} Bokhari, A.H. and Qadir, A.: J. Math. Phys. \textbf{31}(1990)1463.

\item{[3]} Bokhari, A.H. and Qadir, A.: J. Math. Phys. \textbf{34}(1993)3543.

\item{[4]} Ziad, M., Ph.D. Thesis (Quaid-i-Azam University, 1990).

\item{[5]} Qadir, A. and Ziad, M.: Nuovo Cimento \textbf{B110}(1995)317.

\item{[6]} Howard, D. and Stachel, J. (editors): \textit{Einstein and the
            History of General Relativity} (Birkhauser, Boston, 1989).

\item{[7]} Hayashi, K. and Shirafuji, T.: Phys. Rev. \textbf{D19}(1979)3524.

\item{[8]} Weitzenb\"{o}ck, R.: \textit{Invarianten Theorie} (Gronningen: Noordhoft, 1923).

\item{[9]} De Andrade, V.C. and Pereira, J.G.: Phys. Rev. \textbf{D56}(1997)4689.

\item{[10]} Aldrovandi, R. and Pereira, J.G.: {\it An Introduction to
            Gravitation Theory} (preprint).

\item{[11]} M{\o}ller, C.: K. Dan. Vidensk. Selsk. Mat. Fys. Skr. \textbf{1}(1961)10.

\item{[12]} Pellegrini, C. and Plebanski, J.: K. Dan. Vidensk. Selsk.
            Mat. Fys. Skr. \textbf{2}(1962)4.

\item{[13]} Hayashi, K. and Nakano, T.: Prog. Theor. Phys. \textbf{38}(1967)491.

\item{[14]} Hayashi, K.: Phys. Lett. \textbf{B69}(1977)441.

\item{[15]} Hehl, F.W. and Macias, A.: Int. J. Mod. Phys. \textbf{D8}(1999)399.

\item{[16]} Obukhov, Yu N., Vlachynsky, E.J., Esser, W., Tresguerres, R. and Hehl,
            F.W.: Phys. Lett. \textbf{A220}(1996)1.

\item{[17]} Baekler, P., Gurses, M., Hehl, F.W. and McCrea, J.D.: Phys. Lett.
            \textbf{A128}(1988)245.

\item{[18]} Vlachynski, E.J., Esser, W., Tresguerres, R. and Hehl, F.W.: Class.
            Quantum Grav. \textbf{13}(1996)3253.

\item{[19]} Ho, J.K., Chern, D.C. and Nester, J.M.: Chin. J. Phys. \textbf{35}(1997)640.

\item{[20]} Hehl, F.W., Lord, E.A. and Smalley, L.L.: Gen. Relativ. Gravit.
            \textbf{13}(1981)1037.

\item{[21]} Kawai, T. and Toma, N.: Prog. Theor. Phys. \textbf{87}(1992)583.

\item{[22]} Pereira, J.G., Vargas, T. and Zhang, C.M.: Class. Quantum Grav.
            \textbf{18}(2001)833.

\item{[23]} Nashed, G.G.L.: Phys. Rev. \textbf{D66}(2002)064015.

\item{[24]} Kawai, T. and Toma, N.: Prog. Theor. Phys. \textbf{38}(1992)583.

\item{[25]} Sharif, M. and Amir, M. Jamil.: Gen. Relativ. Gravit. \textbf{38}(2006)1735.

\item{[26]} Sharif, M. and Amir, M. Jamil.: Gen. Relativ. Gravit. \textbf{39}(2007)989.

\item{[27]} Sharif, M. and Amir, M. Jamil.: Mod. Phys. Lett. \textbf{A22}(2007)425.

\item{[28]} Sharif, M. and Amir, M. Jamil.: Mod. Phys. Lett. \textbf{A}(2008, to appear).

\item{[29]} Sharif, M. and Amir, M. Jamil.: Mod. Phys. Lett.
            \textbf{A23}(2008)963.

\item{[30]} Stephani, H., Kramer, D., MacCallum, M.A.H., Hoenselaers, C. and
            Hearlt, E.:\textit{ Exact Solutions of Einstein Field Equations}
            (Cambridge University Press, 2003).

\item{[31]} Hartle, J.B.: \emph{ Gravity: An Introduction to Einstein's
            General Relativity} (Baba Barkha Nath Printers, India, 2006).

\item{[32]} Maartens, R. and Maharaj, S.D.: Class. Quantum Grav.
\textbf{3}(1986)1005.

\end{description}
\end{document}